# *STICKERS* EN FACEBOOK: MULTIFUNCIONALIDAD Y CORTESÍA VALORIZANTE EN LA INTERACCIÓN SOCIAL COTIDIANA

Laura M.ª Porrino-Moscoso[1]
*Universidad Alfonso X El Sabio (España)*

Resumen

Los *stickers* o pegatinas son recursos multimodales ampliamente usados en las conversaciones digitales cotidianas. A pesar de su popularidad, la mayoría de los estudios se ha centrado en emojis y emoticonos. Por ello, este trabajo analiza, desde una perspectiva sociopragmática, el uso de *stickers* en los comentarios de un corpus de publicaciones de Facebook con actos de cortesía valorizante, elaboradas durante y después de la pandemia de la COVID-19. El objetivo principal es identificar sus funciones comunicativas y determinar en qué medida actúan como estrategias de cortesía valorizante, considerando también la variable de género. Los resultados muestran que predominan los *stickers* desnudos y aquellos que representan emociones y gestos humanos y situaciones festivas. Se identificaron seis funciones principales: afectiva, ilocutiva, interaccional, gestual, estética y representativa o sustitutiva, y se constató que los *stickers* pueden intensificar mensajes corteses y expresar cortesía valorizante de manera autónoma. Además, se observaron diferencias de género: las mujeres emplean más *stickers*, sobre todo tiernos y afectuosos, mientras que los hombres prefieren figuras humanas masculinas. Estos hallazgos evidencian el papel clave y multifuncional de los *stickers* en la comunicación afectiva digital.

PALABRAS CLAVE: *stickers; emojis; comunicación mediada por ordenador; cortesía valorizante; Facebook*.

Abstract

Stickers are multimodal resources widely used in everyday digital conversations. Despite their popularity, most studies have focused on emojis and emoticons. Therefore, this study analyzes, from a sociopragmatic perspective, the use of stickers in the comments from a corpus of Facebook posts containing acts of face-enhancing politeness, created during and after the COVID-19 pandemic. The main objective is to identify their communicative functions and determine the extent to which they act as strategies of face-enhancing politeness, also considering the gender variable. The results show a predominance of naked stickers and those representing human emotions and gestures, and festive situations. Six main functions were identified: affective, illocutionary, interactional, gestural, aesthetic, and representative or substitutive. It was found that stickers can intensify polite messages and express face-enhancing politeness autonomously. Furthermore, gender differences were observed: women use more stickers, especially cute and affectionate ones, whereas men prefer masculine human figures. These findings highlight the key and multifunctional role of stickers in affective digital communication.

---

[1] N.º ORCID: 0000-0002-2379-3467. Dirección de correo: lporrmos@uax.es.

KEYWORDS: *stickers; emojis; computer-mediated communication; face-enhancing politeness; Facebook.*

0. INTRODUCCIÓN

En la era digital, los usuarios dedican gran parte de su tiempo a interactuar, crear, intercambiar y consumir contenido en redes sociales y servicios de mensajería instantánea. Estos medios sociales han redefinido la comunicación mediada digital, dando lugar a nuevas prácticas discursivas marcadas por la inmediatez, la multimodalidad, la interactividad y la expresividad (Vela Delfa y Cantamutto 2021: 16). Dentro de estas prácticas, sobresale el uso extendido en los discursos digitales de los graficones[2] (Herring y Dainas 2017: 2185), entendidos como elementos gráficos multimodales que abarcan emoticonos :-), *kaomojis* o( "^ ▾ ^" )o, emojis 😄, *stickers*, *GIF*, avatares, memes, imágenes y vídeos.

En la actualidad, la popularidad de los emojis está siendo eclipsada por los *stickers* (en español, *pegatinas*[3]), ilustraciones o animaciones de mayor tamaño y complejidad visual, introducidas en 2011 por la aplicación de mensajería instantánea japonesa LINE y difundidas con celeridad a otras plataformas sociales. Catalogados como los «nuevos emojis» (Jezouit 2017), los *stickers* triunfan en Japón, China, Taiwán, Vietnam, Corea del Sur e Indonesia (Jessica y Franzia 2017: 291; Herring 2018: 10), donde su diseño y comercialización han dado lugar a una industria emergente. De igual modo, investigaciones en español (Vela Delfa y Cantamutto 2021: 28; Sampietro 2023a: 281; Pérez-Sabater 2025: 22) señalan que los jóvenes tienden a preferir los *stickers* frente a los emojis para mantenerse a la vanguardia y desvincularse del uso abundante de emojis de sus padres.

---

[2] En inglés, *graphicon* es la combinación de los términos *graphical* ('gráfico') e *icon* ('icono') (Herring y Dainas 2017: 2185).
[3] Dado su amplio empleo tanto en los medios sociales como en la investigación sobre comunicación mediada por ordenador, en el presente trabajo se opta por utilizar el término *sticker* en lugar de *pegatina* o de sus adaptaciones *estíquer* y *estíker* (Fundéu 2017).

A pesar de su éxito, resulta llamativa la escasez de estudios sobre los *stickers* en comparación con la abundante literatura dedicada a los emoticonos y los emojis. En general, estos trabajos los caracterizan como recursos multimodales (Herring y Dainas 2017; Tang y Hew 2019; Sampietro 2023b), exploran las motivaciones que influyen en su uso y sus funciones comunicativas (Zhou *et al.* 2017; Konrad *et al.* 2020; Tang *et al.* 2021; Oberwinkler 2023), y reflexionan sobre su impacto en el cambio lingüístico y cultural (Konrad *et al.* 2020; Cantamutto 2023; Sampietro 2023a). Asimismo, la mayoría se han centrado en analizar su empleo en las interacciones a través de las aplicaciones de mensajería LINE, WhatsApp y Messenger, pero apenas existen estudios sobre cómo se utilizan en los comentarios de las publicaciones de Facebook, y los pocos disponibles se han enfocado en grupos públicos dedicados a graficones (Herring y Dainas 2017).

Por ello, la presente investigación tiene como finalidad examinar desde un punto de vista sociopragmático los *stickers* en las conversaciones digitales que emergen en los comentarios de Facebook. Conocida como «la red de la amistad», esta plataforma es el medio social con más usuarios en el mundo (3070 millones) y el tercero (20,4 millones) en España, solo por detrás de WhatsApp e Instagram (DataReportal 2025; Kemp 2025). Nuestro estudio analiza un corpus de comentarios de Facebook escritos entre marzo de 2020 y junio de 2023, durante la pandemia y la pospandemia de la COVID-19, en respuesta a publicaciones que realizan actos de habla expresivos, como felicitaciones, agradecimientos y buenos deseos, los cuales suelen propiciar comentarios similares que refuerzan las conexiones sociales. El objetivo es identificar las funciones comunicativas de los *stickers* del corpus y determinar si estos recursos operan como estrategias de cortesía valorizante. Para ello, se analizará también la variable de género. Las preguntas que guían nuestra investigación son las siguientes:

0. ¿Qué tipos de *stickers* predominan en Facebook?
1. ¿Qué funciones comunicativas cumplen?
2. ¿En qué medida sirven para expresar cortesía valorizante?
3. ¿Existen diferencias de género en su uso?

El artículo se organiza de la siguiente manera: tras esta introducción, se presenta el marco teórico que sustenta nuestro trabajo, centrado en la comunicación mediada multimodal, la conceptualización integral de los *stickers* y la teoría de la cortesía valorizante. A continuación, se detalla la metodología empleada, seguida del análisis y de los resultados obtenidos. Finalmente, se exponen las conclusiones derivadas de la investigación.

1. Marco teórico

1.1 *Los graficones en la comunicación mediada multimodal*

La comunicación mediada por ordenador (CMC) comprende los intercambios realizados mediante tecnologías digitales (Herring 1996a: 1) y se sitúa en un *continuum* entre oralidad y escrituralidad (Koch y Oesterreicher 2007: 30; Crystal 2011: 19). La naturaleza social de las redes sociales y de la mensajería instantánea impulsa a los internautas a reproducir en sus interacciones en línea rasgos propios de la conversación cara a cara, como la inmediatez, la espontaneidad, la brevedad, la coloquialidad y el uso de elementos no verbales. A ello se suma la multimodalidad inherente a estos medios, en los que convergen múltiples modos semióticos (Adami y Jewitt 2016: 264; Jewitt *et al.* 2016: 25; Bateman *et al.* 2017: 360; Hasym y Arafah 2023: 98; Sampietro 2023b: 9; Vela Delfa y Cantamutto 2024: 784). En estos entornos, textos, imágenes, vídeos, audios, emoticonos, *kaomojis*, emojis, *stickers*, avatares, GIF, memes, *hashtags* o enlaces actúan como constructores de sentido del discurso (Kress y van Leeuwen 2001: 3) y convierten a los usuarios en prosumidores mediáticos con autonomía creativa. Esta dinámica exige una alfabetización multimodal (Bateman 2008: 2), necesaria para interpretar la variedad de graficones, recursos gráficos que aportan expresividad y ejercen como pistas contextualizadoras que orientan la interpretación de los mensajes (Gumperz 1982).

Los graficones, ilustrados en la figura 1, surgieron como respuesta a la falta de señales no verbales en la comunicación digital. Los primeros fueron los emoticonos[4] con caracteres ASCII :-) y :-( propuestos por Scott Fahlman en 1982 como marcadores de broma y de seriedad, respectivamente (McCulloch 2019: 261). En Japón aparecieron simultáneamente los *kaomojis*, centrados en los ojos y con mayor carga expresiva (Katsuno y Yano 2002: 206; Moschini 2016: 21). No obstante, los graficones de mayor alcance global son los emojis[5], empleados por el 92 % de los internautas (Unicode 2024). Aunque su creación suele atribuirse a Shigetaka Kurita, diseñador en 1999 de 176 emojis para el servicio móvil NTT DoCoMo, el operador J-Phone ya había lanzado 90 emojis en 1997. Su expansión se consolidó con la estandarización de Unicode impulsada por Google y su incorporación al teclado de Apple en 2011 (Vela Delfa y Cantamutto 2021: 25).

---

[4] Por aquel entonces, solía denominarse *smiley* ('carita sonriente'), pero pronto comenzó a generalizarse el acrónimo inglés *emoticon* ('emoticono'), formado a partir de la unión de *emotion* ('emoción') e *icon* ('icono') (Halté 2019: 370).

[5] El término *kaomoji* deriva de *kao* ('cara') y *moji* ('carácter' o 'letra'), y significa 'carácter facial' (McCulloch 2019: 262). Por su parte, *emoji* resulta de la combinación de *e* ('imagen') y *moji* ('carácter' o 'letra') (Konrad *et al.* 2020: 221). La similitud con emoticono es meramente accidental.

FIGURA 1. *Ejemplos de graficones en la comunicación digital*

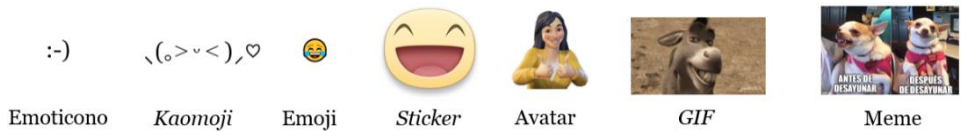

| Emoticono | *Kaomoji* | Emoji | *Sticker* | Avatar | *GIF* | Meme |

En 2011, la empresa surcoreana Naver diseñó para la aplicación de mensajería japonesa LINE los *stickers*, ilustraciones o animaciones de personajes de mayor tamaño, complejidad y expresividad que los emojis. Su diseño más elaborado y centrado en personajes, que a menudo muestra el cuerpo completo, así como su mayor detallismo visual, permiten reproducir con más fidelidad el lenguaje corporal, las emociones, las actitudes y las acciones (S. Kato y Y. Kato 2018: 4; Konrad *et al.* 2020: 222). Dado su potencial, fueron incorporados en 2013 en Messenger y, al año siguiente, en los comentarios de Facebook, así como en otros medios como WhatsApp, Telegram, Snapchat, Instagram, o TikTok. Los *GIF* (pequeñas secuencias de vídeo en bucle) y los memes (montajes de texto e imagen que suelen viralizarse en medios sociales) se han consolidado como símbolos culturales de internet (McCulloch 2019: 351). Recientemente, se han implementado avatares personalizados en formato de *stickers*, ampliando las posibilidades expresivas de la comunicación digital.

### 1.2 *El* boom *de los* stickers: *caracterización, tipología y presencia en Facebook*

Los *stickers*, también denominados *pegatinas*, *calcomanías* o *sellos* (*sutanpu* en japonés), son imágenes coloridas o animaciones, amplias y expresivas, con estética de dibujos animados, a las que puede añadirse o no texto. Se ofrecen como recursos organizados en paquetes personalizados y coleccionables (de Seta 2018; Konrad *et al.* 2020: 222). Se consideran una evolución de los emoticonos y emojis, por lo cual Liu y Sun (2020: 9) los describen como «emojis de nueva generación». Desarrollados en 2011 para la aplicación LINE, los *stickers* se inspiran en la cultura visual japonesa, especialmente en los personajes del manga, el anime, los videojuegos y la estética *kawaii* ('tierna' o 'bonita') (Steinberg 2020: 3).

LINE permitió que cualquier usuario diseñara, compartiera o vendiera sus *stickers* en la aplicación, lo que propició el surgimiento de una potente industria, en la que los diseñadores se convirtieron en productores culturales (Steinberg 2020: 1). Este modelo se expandió a otras plataformas orientales y occidentales, que incorporaron tiendas, bibliotecas gratuitas e incluso sistemas de recomendación de *stickers*. En Messenger, por ejemplo, en 2018 los internautas enviaron más de 380 millones de *stickers* al día, frente a los 22 millones de *GIF* (Susanto 2018).

A diferencia de los emojis, regulados por Unicode, los *stickers* son una clase abierta (Vela Delfa y Cantamutto 2021: 28), lo que facilita la comercialización de colecciones inspiradas en personajes de anime, manga, series, películas, libros, celebridades o eventos, es decir, contenidos que nunca podrían integrarse como emojis por su carácter abierto, personalizado y no estandarizado (Porrino-Moscoso 2024: 154). También es habitual que empresas y marcas diseñen sus propios *stickers* con fines promocionales (Steinberg 2020: 5; Kang *et al.* 2022: 2). Además, los usuarios recurren a aplicaciones como Sticker.ly o Sticker Maker para crear *stickers* creativos, y cada vez más medios sociales permiten generarlos directamente a partir de imágenes, dibujos y textos, e incluso mediante inteligencia artificial.

Hoy en día, los *stickers* son omnipresentes en las interacciones socioemocionales en línea (Derks *et al.* 2007: 842). Según Konrad *et al.* (2020: 217), los *stickers* se sitúan en una fase alta, con un uso creciente y una mayor marcación pragmática, entendida como la propiedad de destacar frente a formas más comunes y asociada a una mayor intensidad emocional, positividad, intimidad y carácter lúdico. Por el contrario, los emojis se encuentran en una fase de convencionalización y desmarcación pragmática, en la que tienden a volverse más rutinarios, a atenuar su carga emocional y pueden emplearse en mensajes semiserios y en relaciones menos cercanas. En el este asiático, los *stickers* han comenzado a desplazar a los emojis, en parte por la complejidad tipográfica de sus sistemas de escritura (Ma 2016: 19)[6]. Asimismo, pueden entenderse como signos no verbales demostrativos (Konrad *et al.* 2020: 222). Al igual que otros graficones, se utilizan de modo predominantemente voluntario, como signos ostensivos más que inconscientes (Derks *et al.* 2008: 778).

Debido a su tamaño, estos pictogramas no se insertan en la misma línea de texto, sino que aparecen como mensajes individuales[7]. Esta disposición les confiere una notable autonomía comunicativa, pues pueden comunicar un concepto, una acción o incluso una historia, sin apoyo textual, y funcionar como actos ilocutivos completos (Jessica y Franzia 2017: 292; Konrad *et al.* 2020: 229; Tang *et al.* 2021: 10). Algunos *stickers* incluyen únicamente texto y producen el mismo efecto que un mensaje escrito (Zhou *et al.* 2017: 756; S. Kato y Y. Kato 2018: 14). Asimismo, las colecciones de *stickers* suelen ser exclusivas de cada servicio y carecen de interoperabilidad con otros medios.

---

[6] En España, los emojis y los emoticonos fueron elegidos «palabra del año» por la Fundación del Español Urgente (FundéuRAE) en 2019. Sin embargo, el término *sticker* aún no figura en el *Diccionario de la lengua española* de la Real Academia Española, y la entrada *pegatina* se limita a su acepción física, sin reflejar su uso digital. En cambio, el *Diccionario de Americanismos* de la Asociación de Academias de la Lengua Española sí incorpora la adaptación *estíquer*.

[7] En octubre de 2025, Instagram introdujo la posibilidad de añadir *stickers* en los mensajes privados, los cuales pueden ubicarse de forma independiente en cualquier parte del chat, sin necesidad de integrarse en una línea de texto (Moreno 2025).

En cuanto a su tipología, Tang *et al.* (2021: 9-10) diferencian entre seis tipos según la presencia de texto y su dinamismo: estáticos con solo texto, estáticos con solo imagen, estáticos con texto e imagen, animados con solo texto, animados con solo imagen y animados con texto e imagen. Por su temática, pueden agruparse en categorías como emociones y gestos humanos, animales, personajes de ficción o celebridades, interacciones sociales, acciones y objetos cotidianos, festividades y celebraciones, humor o elementos decorativos (estrellas, globos, etc.). Destacan los *stickers* antropomórficos y los humorísticos, usualmente en formato de meme (Sampietro 2023b: 13; Alvarado Ortega y Linares Bernabéu 2024: 19).

En la macroplataforma Facebook, los internautas pueden añadir *stickers* a sus propias publicaciones o, más habitualmente, incorporarlos en los comentarios a publicaciones ajenas. Para ello, seleccionan estos elementos de sus colecciones personales, integradas en la amplia biblioteca de *stickers* de la red, organizada por diversas categorías afectivas (feliz, triste, enojado, enamorado, activo, soñoliento y confundido) y tipos de acción (celebrando, trabajando y comiendo). Un buscador interno facilita su localización por palabras clave, y la tienda ofrece numerosos paquetes temáticos gratuitos, que suelen incluir entre 16 y 20 *stickers* cada uno. En total, se han contabilizado 131 paquetes disponibles, la mayoría creados por diseñadores y estudios de animación reconocidos, como Quan Inc., Ghostbot, Minto, Funnyeve y David Lanham. También se incluyen paquetes propios de Facebook y Messenger, así como colecciones basadas en franquicias populares, como Bob Esponja o Snoopy y Brown, y personajes de *webseries* y tiras cómicas, como Los amigos de Tontón, Pusheen o Tuzki.

1.3 *Motivaciones de uso, factores y funciones comunicativas de los* stickers

Las investigaciones que han examinado las motivaciones de uso de los *stickers* evidencian que estos elementos son útiles, precisos, bonitos, disfrutables y fáciles de usar. Se emplean porque propician la autoexpresión y la comunicación emocional, dinamizan las interacciones, aportan humor y ternura en los mensajes, y contribuyen a la gestión de la identidad personal y social en los entornos digitales (W. Lee y Y.-H. Lin 2019: 51; Konrad *et al.* 2020: 224-225; Kang *et al.* 2022: 1). Asimismo, actúan como marcadores de pertenencia grupal, fortalecen la cohesión social, y resultan muy prácticos en situaciones en las que no se sabe qué responder (Tang *et al.* 2021: 585). Algunos usuarios manifiestan sentirse identificados con los *stickers* y disfrutan eligiendo aquellos que mejor ilustran sus sentimientos o que consideran acordes con los gustos o el parecido físico de sus interlocutores (Zhou *et al.* 2017: 753).

Por otro lado, el uso y la interpretación de los *stickers* dependen de factores personales, contextuales, sociales y tecnológicos (J. Lee *et al.* 2016: 765; Moschini 2016: 21; Tang y Hew 2018: 199; Liu y Sun 2020: 2; Bae *et al.* 2024: 22). Entre ellos, sobresalen los factores contextuales, ya que la comprensión semántica y pragmática de los mensajes con *stickers* varía según la situación comunicativa, el contexto sociocultural, la intención y tono del emisor, el cotexto verbal, la finalidad social o de tarea del intercambio, la temática y el grado de sincronicidad de la interacción. También resulta esencial la relación interpersonal entre los interlocutores, el nivel de confianza y cercanía, la frecuencia de contacto, los gustos y preferencias personales del destinatario y las convenciones sociales.

Con respecto al género, las investigaciones, centradas especialmente en contextos lingüísticos-culturales anglófonos, muestran que las mujeres utilizan emoticonos y emojis con mayor frecuencia que los hombres (Tossell *et al.* 2012; Chen *et al.* 2018; Herring y Dainas 2018, 2020), lo que se ha relacionado con una mayor orientación del género femenino hacia la expresión emocional y relacional, como ya señaló Holmes (1995: 193) en su estudio sobre la interacción conversacional en Nueva Zelanda. No obstante, ambos géneros difieren en la selección de signos y en sus funciones pragmáticas (Herring y Dainas 2020: 2; Sadia y Hussain 2023: 122). De manera similar, los escasos estudios que abordan el género en relación a los *stickers*, en este caso enfocados principalmente en entornos lingüísticos-culturales del este asiático, como el japonés y el chino, indican un mayor uso femenino, sobre todo de aquellos con estética tierna, temática animal y texto integrado (Y. Kato 2017; Oberwinkler 2023: 256; Yang *et al.* 2023: 835).

Los *stickers* son *graficones* multifuncionales capaces de desempeñar simultáneamente múltiples funciones comunicativas. Estos recursos pueden complementar el texto y sustituir palabras y frases. Como los emojis, emoticonos y *kaomojis*, una de sus funciones principales es la expresión de emociones diversas de manera intensificada, como la felicidad, la tristeza, el miedo, la vergüenza o la ira (Tang *et al.* 2021: 592-593; Oberwinkler 2023: 256; Sadia y Hussain 2023: 123). Asimismo, representan con notable precisión gestos faciales y corporales (Sampietro 2023b: 50).

Del mismo modo, desde un punto de vista pragmático, en redes sociales y mensajería instantánea, pueden actuar como actos de habla autónomos (saludos y despedidas, felicitaciones, agradecimientos, elogios, disculpas, etc.); indicar, reforzar o atenuar la fuerza ilocutiva del mensaje; vehicular ironía, sarcasmo o humor; y transmitir (des)cortesía o afecto (Tang *et al.* 2021: 592; Oberwinkler 2023: 256; Sampietro 2023b: 54-55). Por otra parte, contribuyen a la gestión interaccional al dinamizar la conversación, marcar la apertura y el cierre de la misma, señalar la escucha, cambiar el tema y prevenir silencios (J. Lee *et al.* 2016: 763; S. Kato y Y. Kato 2018: 14; Konrad *et al.* 2020: 225; Sampietro 2023b: 48). Además, se pueden usar con fines estéticos o decorativos (Dainas y Herring 2021: 117) y, dada su alta expresividad, ayudan a evitar malentendidos (Tang y Hew 2019: 2467). A partir de estas propuestas, el análisis del corpus (sección 3.2) explora las funciones comunicativas que los *stickers* generan en Facebook.

### 1.4 *La cortesía valorizante en la conversación*

La cortesía actúa como un mecanismo regulador de la interacción humana que favorece la convivencia armoniosa y fortalece los lazos sociales. Este fenómeno sociocultural, cuya manifestación varía según las normas y valores propios de cada comunidad (Wierzbicka 1991: 161), se relaciona con formas de comportamiento consideradas educadas, amables y respetuosas. Desde una perspectiva pragmática, se entiende como un conjunto de estrategias conversacionales orientadas a regular las interacciones, así como a prevenir, mitigar o reparar posibles conflictos (Escandell Vidal 1996: 145).

En la teoría de la cortesía de Brown y Levinson (1987: 61), basada en la noción de imagen (*face* en inglés) de Goffman (1967: 5), la imagen pública se concibe como un bien simbólico y vulnerable, que puede perderse, mantenerse o mejorarse durante la conversación, especialmente ante la presencia de actos amenazadores de la imagen. Para contrarrestar estos riesgos, los hablantes cooperan y aplican actividades de imagen (*face-work*), concebidas como estrategias de cortesía destinadas a preservar la imagen propia y ajena. Como contrapartida a este modelo centrado esencialmente en los aspectos negativos de la interacción, Kerbrat-Orecchioni (1992: 176, 1996: 54) pone en valor la existencia de actos halagadores de la imagen, como los cumplidos, los agradecimientos o los deseos. Estos actos expresivos, en el sentido de la clasificación de Searle (1976: 12), enaltecen la imagen del otro y refuerzan la solidaridad y la afinidad entre los interlocutores. La autora distingue entre cortesía negativa (o mitigadora), que busca evitar o suavizar actos amenazadores de la imagen, y cortesía positiva (o valorizante), que conlleva el uso de actos halagadores de la imagen, preferiblemente reforzados (Kerbrat-Orecchioni 1996: 55-59; Carrasco Santana 1999: 22; Albelda Marco 2003: 300). En culturas de acercamiento como la española (Haverkate 1994: 56), predominan las estrategias de cortesía valorizante y el uso de actos expresivos, tanto en conversaciones presenciales como en línea.

Asimismo, en el ámbito hispánico, conviene señalar la distinción que establece Bravo (1999: 9) entre la imagen de autonomía y afiliación como dos necesidades humanas fundamentales. La imagen de autonomía implica percibirse a uno mismo y ser percibido como miembro diferenciado del grupo, mientras que la imagen de afiliación supone percibirse y ser considerado como parte integrante de este. Se trata de categorías vacías, que se concretan mediante los contenidos de imagen propios de cada comunidad.

Así, en España, la autoafirmación y el orgullo se asocian con la imagen de autonomía, y la confianza y la expresión de afectividad se vinculan con la de afiliación (Bravo 1999: 160, 2002: 145; Hernández Flores 2002: 92).

Por otra parte, investigadores como Bravo (2002: 141), Hernández Flores (2002: 55-56, 2013: 176) y Bernal Linnersand (2007: 37-39) indican que el concepto de «actividad de imagen» es más amplio y comprende tanto comportamientos corteses como no corteses; es decir, va más allá de las actividades de cortesía y no se limita a la gestión de amenazas a la imagen. A este respecto, Hernández Flores (2013: 182) plantea que la combinación de los planos de direccionalidad (dirección del efecto social hacia la imagen del hablante) y de modalidad (tipo de efecto positivo, negativo o neutro en la imagen social) da lugar a tres tipos de actividades de imagen, también observables en los entornos sociales: actividades de cortesía, con un efecto positivo en la imagen del hablante y del destinatario (p. ej. un agradecimiento a una felicitación de cumpleaños); actividades de descortesía, con un efecto negativo en la imagen del oyente y un efecto negativo o positivo en el hablante (p. ej. una crítica insultante a un político puede tener un efecto negativo en la imagen del hablante, pero positivo si hay individuos que lo apoyan); y actividades de autoimagen, con un efecto positivo en la propia imagen del hablante cuando la realza, protege o repara (p. ej. una publicación con una foto para presumir de un logro académico).

En los medios sociales, los emoticonos, emojis, *kaomojis*, *stickers* y *GIF* son recursos multimodales para expresar cortesía y gestionar la imagen interpersonal, de modo que favorecen la protección y ensalzamiento de la propia imagen y la del destinatario (Dresner y Herring 2010: 257; Kaneyasu 2022: 153; Vela Delfa y Cantamutto 2025: 177). En el caso de los *stickers*, la cortesía representa una de sus funciones comunicativas principales (Ikuta 2022: 61; Oberwinkler 2023: 254; Sampietro 2023b: 56). Estos graficones pueden atenuar actos amenazadores de la imagen, tales como peticiones u órdenes, aunque, sobre todo, se ha comprobado que suelen utilizarse como estrategias de cortesía valorizante para reforzar actos halagadores de la imagen, como agradecimientos, cumplidos o deseos (Oberwinkler 2023: 254; Porrino-Moscoso 2024: 834). También ayudan a cerrar de forma cortés la conversación (Tang *et al.* 2021: 596) y pueden transmitir cortesía por sí mismos o intensificar la cortesía expresada verbalmente (Ikuta 2022: 61). A su vez, pueden emplearse con fines descorteses, cuando se usan de manera irónica o agresiva para amenazar la imagen del otro. Por último, Lee *et al.* (2016: 763-764) señalan que los internautas usan *stickers* para la autorrepresentación, el mantenimiento del *statu quo*, la presencial social, la generación de simpatía y la gestión de impresiones, es decir, para proyectar una imagen particular ante los demás, como la de ser una persona cortés.

2. Metodología

El presente estudio se enmarca en la teoría de la comunicación mediada por ordenador (Herring 1996a, 1999), el modelo de cortesía valorizante de Kerbrat-Orecchioni (1992, 1996) y la tipología de actos de habla de Searle (1976). Se adopta un enfoque sociopragmático para analizar los *stickers*, dado que no solo se consideran los rasgos lingüísticos y multimodales, sino también el contexto comunicativo, los factores sociales que condicionan la interacción y la variable de género. La investigación aplica una metodología mixta, que combina el análisis cuantitativo y cualitativo de los discursos.

El corpus está compuesto por publicaciones de 24 usuarios españoles en Facebook, junto con los comentarios que recibieron, y abarca el período de marzo de 2020 a junio de 2023, coincidente con la pandemia y la pospandemia de la COVID-19 (Porrino-Moscoso 2024). Su compilación siguió las directrices de la lingüística de corpus (Rojo Sánchez 2021), complementadas con la observación directa propia de la etnografía virtual (Hine 2000). Se recopilaron publicaciones con actos expresivos de cortesía valorizante: felicitaciones, buenos deseos, agradecimientos, muestras de sentimientos positivos, condolencias, recuerdos y contenidos de presumir de imagen. En este último caso, los sujetos realizaban actividades de imagen (Hernández Flores 2013: 182), como cambiar su foto de perfil o de portada, o compartir una imagen personal, lo que generaba respuestas elogiosas.

La muestra es equilibrada en cuanto al género (masculino y femenino) y a la generación etaria (*baby boomers*, generación X, generación Y o *millenials*, y generación Z)[8], y fue anonimizada conforme a los principios éticos de la CMC (Herring 1996b; Beiβwenger y Storrer 2008). Los usuarios incluidos son amistades de Facebook de la autora, dado que era necesario conocer sus relaciones interpersonales para comprender cómo conversan y se comportan con sus lazos fuertes. Previamente, la autora había solicitado en su perfil de Facebook la participación voluntaria en esta investigación. De los 42 contactos que respondieron, se aplicó la técnica de muestreo por cuotas de afijación simple, con el fin de evitar el azar y posibles desequilibrios en la proporción de hombres y mujeres en los distintos grupos de edad. Esta técnica consiste en dividir la población de estudio en subpoblaciones para garantizar la representación de todas ellas (López Morales 1994: 58-59). Los usuarios seleccionados debían ser españoles, escribir en español, ser activos en la publicación en Facebook y recibir numerosas gratificaciones (Me gustas, reacciones, comentarios y comparticiones).

Por otra parte, de los 24 usuarios seleccionados, se examinaron los diez primeros comentarios de cada publicación, realizados por familiares, amistades y compañeros de trabajo o estudios de los autores, junto con las respuestas de estos últimos. Los *stickers* examinados pertenecen a los comentarios de esos usuarios y a las réplicas de los autores. El corpus contiene 312 publicaciones y 3148 comentarios. Fue compilado y examinado con el programa de análisis de datos cualitativos Atlas.ti (versión 8), que permitió organizar los datos y facilitar el proceso de análisis. Los *stickers* fueron codificados manualmente y clasificados según el paquete temático al que pertenecían, su función en la interacción y el remitente (autor o comentador). Para garantizar la consistencia de la codificación, se realizó de manera iterativa, revisando y ajustando las categorías a lo largo del análisis.

---

[8] Se adoptan los puntos de corte generacionales establecidos por el Pew New Research (Dimock 2019).

3. ANÁLISIS Y RESULTADOS DE LOS COMENTARIOS CON *STICKERS*

3.1 *Tipos de* stickers *del corpus*

En el corpus de publicaciones de Facebook se ha contabilizado un total de 338 *stickers*: 319 proceden de comentarios realizados por familiares, amistades y compañeros de trabajo o de estudios, y 19 pertenecen a las respuestas de los propios autores de las publicaciones. Con respecto al género, la gran mayoría de los *stickers* enviados por los comentadores, el 86,2 %, fueron publicados por mujeres y solo el 13,8 % por hombres. De manera similar, entre los autores, las mujeres compartieron el 84,2 % de los *stickers*, mientras que los hombres el 15,8 %. Entre los comentadores, este predominio femenino se explica, en parte, por el mayor número de comentarios producidos por mujeres en el corpus, lo que resulta coherente con estudios previos (Kimbrough *et al.* 2013: 898; Herring y Stoerger 2014: 569) que muestran que, en redes sociales como Facebook, Twitter (actualmente X) o Pinterest, las mujeres suelen mostrarse más participativas y comprometidas en las interacciones digitales. Además, como se ha observado en Facebook (Herring y Dainas 2017, 2020), LINE (Y. Kato 2017; Oberwinkler 2023), WhatsApp (Pérez-Sabater 2019; T. K. Koch *et al.* 2022), WeChat (Yang *et al.* 2023), en los mensajes de iPhone (Tossell *et al.* 2012) y en teclados virtuales (Chen *et al.* 2018), y como se aprecia también en el corpus, las mujeres tienden a ser más expresivas y usan con más asiduidad *stickers*, emojis o emoticonos para transmitir la carga afectiva de sus mensajes.

La Tabla 1 evidencia que los *stickers* enviados por mujeres (autoras y comentadoras) se dirigieron principalmente a otras mujeres (51,5 %), pero también a hombres (31,7 %). En cambio, los hombres (autores y comentadores) mandaron más a menudo estos graficones a otros hombres (9,5 %) que a mujeres (4,1 %). Una proporción reducida (3,3 %) de *stickers* tuvo un destinatario colectivo, al estar dirigida a todos los amigos del perfil.

Tabla 1. *Distribución de los stickers según el género del emisor y del destinatario*

| Género del emisor | A hombres | A mujeres | A los amigos | Total |
|---|---|---|---|---|
| Hombres | 32 (9,5 %) | 14 (4,1 %) | 2 (0,6 %) | 48 (14,2 %) |
| Mujeres | 107 (31,7 %) | 174 (51,5 %) | 9 (2,7 %) | 290 (85,8 %) |
| **Total** | 139 (41,1 %) | 188 (55,6 %) | 11 (3,3 %) | 338 (100 %) |

Al compartir un *sticker*, los usuarios pueden mandarlo junto con un mensaje de texto, en cuyo caso aparece en la línea siguiente, como se ilustra en (1)[9], o bien enviarlo de forma aislada, como se muestra en (2). En este último caso, el *sticker* actúa como un *sticker* desnudo, análogo al emoticono desnudo sin acompañamiento textual, de acuerdo con la propuesta de Provine (2007: 207). También es habitual que publiquen primero un comentario de texto y, a continuación, otro comentario compuesto únicamente por un *sticker*.

(1) Comentadora (felicitar aniversario): Muchas felicidades y a seguir cumpliendo !!

(2) Comentador (mostrar sentimiento positivo):

En las muestras conjuntas de autores y comentadores, se ha constatado que el 66,9 % de los *stickers* se presentan desnudos, frente al 33,1 % que se envían acompañados de un mensaje de texto. Además, se han identificado *stickers* combinados con emojis, con o sin texto, o con menciones hipervinculadas al autor o a otros comentadores para llamar su atención. Asimismo, el 24,9 % de los *stickers* del corpus son textuales, es decir, incorporan palabras o expresiones, y una pequeña parte incluye números, como en (3), donde el comentador utiliza el *sticker* «2021» junto a la palabra «Feliz» para desear un Feliz Año Nuevo.

(3) Comentadora (desear Feliz Año): Feliz

---

[9] Los ejemplos han sido extraídos del corpus y se ha conservado su ortografía y tipografía originales, incluidos los errores, dada la información relevante que aportan. En cada ejemplo, se indica el tipo de acto de habla de la publicación de la que se extrajo, el género del participante y si el *sticker* fue enviado por un comentador o por un autor.

Resulta relevante que el 60 % de estos *stickers* están en inglés, y se usan, especialmente, para felicitar el cumpleaños (*Happy Birthday*), elogiar a alguien (*SUPER!*), desear un Feliz Año Nuevo (*HAPPY NEW YEAR!*) o una Feliz Navidad (*Merry Christmas*), como se aprecia en el ejemplo (4). Estos *stickers* son claros, visuales y muy habituales en festividades y celebraciones.

(4) Autora (desear Feliz Navidad): Muchas gracias primores🥰🥰🥰

De hecho, solo el 38,7 % de los *stickers* son en español, y se emplean principalmente para agradecer (*GRACIAS*), felicitar o formular un cumplido (*BRAVO!*), expresar emociones (*Qué Orgullo*) o desear un Feliz Día del Padre (*FELIZ DÍA DEL Padre*), como en (5). También se halló un *sticker* en forma de corazón con un mensaje mixto en inglés y español (*SORRY Hola Te Amo Adiós*). Independientemente del idioma, estos *stickers* textuales suelen aparecer en mayúsculas. La media de palabras por *sticker* es de aproximadamente 1,6 en español y 1,9 en inglés.

(5) Autor (desear Feliz Día del Padre):

Como se ilustra en la Tabla 2, a nivel temático priman los *stickers* que representan emociones y gestos humanos (36,1 %), seguidos por los de festividades y celebraciones (17,2 %), que incluyen todos aquellos, con o sin texto, relacionados con fechas especiales, como cumpleaños o Navidad. También son recurrentes los *stickers* usados en interacciones sociales (13 %), por ejemplo, para agradecer (*gracias*) o expresar cortesía (*de nada*). De igual modo, destacan los *stickers* de personajes de ficción reconocidos (Snoopy, Bigli y Migli, Pucca, BT21, etc.) (16,6 %) y los *stickers* de animales (9,2 %), normalmente antropomorfizados, que simulan emociones y expresiones faciales o corporales humanas. Los *stickers* de animales y algunos de personajes de ficción tienen una estética marcadamente *kawaii*, la cual, como señala Oberwinkler (2023: 245), denota una mayor ternura respecto a los *stickers* humanos. Los perros, las nutrias, los conejos y los osos fueron muy frecuentes. En menor medida, se localizaron *stickers* de acciones y objetos cotidianos (7,4 %). Solo se detectaron dos *stickers* con efecto humorístico (0,6 %), ambos con representaciones de perros, mientras que no se registró ningún *sticker* perteneciente a la categoría de elementos puramente decorativos (estrellas, globos, etc.).

Tabla 2. *Frecuencia de las categorías temáticas de stickers del corpus*

| Categoría temática | Frecuencia | Porcentaje |
|---|---|---|
| Humanos | 122 | 36,1 % |
| Animales | 31 | 9,2 % |
| Personajes de ficción reconocidos | 56 | 16,6 % |
| Acciones y objetos cotidianos | 25 | 7,4 % |
| Interacciones sociales | 44 | 13 % |
| Festividades y celebraciones | 58 | 17,2 % |
| Humor | 2 | 0,6 % |
| Elementos decorativos | 0 | 0 % |
| **Total** | 338 | 100 % |

La Tabla 3 muestra las colecciones de *stickers* más frecuentes y los *stickers* más empleados. Cabe señalar que *SUPER!* y *BRAVO!* no se pudieron asociar a ninguna categoría temática, y se consideran *stickers* sueltos.

Tabla 3. *Colecciones de stickers más frecuentes y stickers más empleados*

| Colecciones de *stickers* | | | | | | | | | |
|---|---|---|---|---|---|---|---|---|---|
| Hacker girl | Valientes y fabulosas | Fiestas en casa | Los amigos de Tonton | SUPER! | Mugsy | Moodies | Daily Duncan | Angry birds | BRAVO! |
| 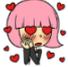 | 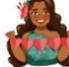 | 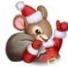 | 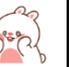 | 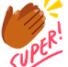 | 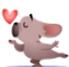 | 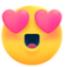 | 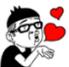 | 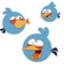 | 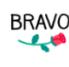 |
| 35 | 24 | 14 | 12 | 12 | 10 | 9 | 9 | 9 | 7 |
| *Stickers* | | | | | | | | | |
| 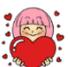 | 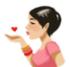 | 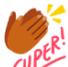 | 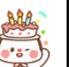 | 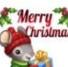 | 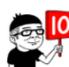 | 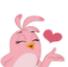 | 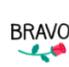 | 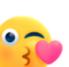 | 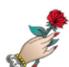 |
| 28 | 18 | 12 | 10 | 9 | 8 | 8 | 7 | 6 | 6 |

En determinadas ocasiones, los comentadores y los autores intercambian *stickers* del mismo tema o emoción, dando lugar a una especie de «guerra de *stickers*». Este intercambio consiste en responder a un *sticker* con otro *sticker* y cumple una función lúdica y relacional, documentada por Konrad *et al.* (2020: 224-225) con usuarios angloparlantes en Messenger y también observada en el corpus de Facebook analizado. Igualmente, Sampietro (2023b: 60) señala que este uso es frecuente en interacciones diádicas o grupales informales en español, y funciona como una reelaboración humorística. Asimismo, se han detectado casos de participantes que recurren de manera reiterada a los mismos *stickers* para determinadas situaciones comunicativas, como desear un Feliz Año Nuevo, expresar agradecimiento o felicitar el cumpleaños. De igual modo, es común que se produzca un efecto de «contagio» (McCulloch 2019: 279), de manera que en una misma publicación varios comentadores utilicen el mismo *sticker*, por ejemplo, los mismos *stickers* navideños. De hecho, Facebook permite guardar y reutilizar los *stickers* empleados por otros usuarios, lo que facilita esta propagación.

3.2 *Funciones comunicativas de los* stickers *del corpus*

La multifuncionalidad de los *stickers* también es apreciable en el corpus, en tanto que un mismo *sticker* puede desempeñar simultáneamente varias funciones comunicativas. Así, en (6), la autora agradece el apoyo recibido a todas las personas que la ayudaron a aprobar la oposición, y una amiga la felicita mediante un *sticker* que contiene la interjección *BRAVO!* acompañada de una rosa. Esta interjección puede considerarse también un cumplido (Placencia y Eslami 2020). De esta manera, este graficón cumple con una función ilocutiva, al realizar directamente el acto de felicitación o cumplido, y una función afectiva, al expresar aprecio y cercanía con la rosa. En consecuencia, refuerza el lazo interpersonal, ya que combina reconocimiento y afecto en una única pieza visual.

(6) Comentadora (felicitar logro):

Teniendo en cuenta la revisión de la literatura sobre las motivaciones de uso y funciones de los *stickers* expuesta en el apartado 1.3, y tomando como referencia principalmente las investigaciones de Sampietro (2023b) sobre la comunicación digital española con distintos graficones, en el análisis del corpus se han identificado las siguientes funciones de los *stickers*: afectiva, ilocutiva, interaccional, gestual, estética y representativa o sustitutiva. Estas funciones se distinguen analíticamente para describir con mayor precisión los efectos que los *stickers* generan en las interacciones en Facebook. Sampietro (2023b: 44) delimita como funciones de los recursos multimodales: indicar contenidos emocionales, emplear secuencias de comunión fática, representar gestos, realizar o especificar actos de habla con posibles implicaciones de cortesía y expresar o responder al humor. Las categorías distinguidas en el corpus se alinean en gran medida con estos usos, al mismo tiempo que se complementan con las funciones adicionales estética y representativa o sustitutiva (Dainas y Herring 2021: 117; Sadia y Hussain 2023: 123), necesarias para dar cuenta de los usos observados.

En primer lugar, la respuesta predominante a los actos expresivos de cortesía valorizante de las publicaciones es la función afectiva, ya que permite la expresión de emociones y afectos, lo que favorece el fortalecimiento de los vínculos sociales. Siguiendo la clasificación de Searle (1976: 12-13), estos actos se pueden concebir como actos expresivos, dado que comunican estados psicológicos o actitudes del emisor hacia los demás. Esta función comprende, a su vez, dos dimensiones complementarias: la emotiva y la afectuosa. Por un lado, y al igual que en la investigación de Tang *et al.* (2021: 592-593) sobre internautas chinos que utilizan la aplicación de mensajería WeChat, se observan *stickers* que representan una amplia variedad de sentimientos y emociones, concretamente alegría, tristeza, vergüenza, sorpresa y orgullo. En (7), la amiga reacciona a la publicación de la autora ofreciendo su pésame, transmitiendo su afecto, empatía y apoyo, y reforzando su expresión de tristeza con emojis de llanto fuerte. Además, en una segunda intervención, usa un *sticker* de llanto intenso, con el que reafirma la solidaridad y la cercanía hacia la autora.

(7) Comentadora (expresar condolencias): lo siento desde lo más profundo de mi corazón de verdad […] muchos besos a tu madre y vosotras de mi parte corazon
Comentadora:

Por otro lado, la mayor parte de los *stickers* de las muestras examinadas expresan afectuosidad, en concreto, distintas manifestaciones de aprecio. La afectuosidad, percibida como una de las manifestaciones afectivas más necesarias del ser humano y como una de las características de la imagen de afiliación española (Hernández Flores 2002: 92), se refleja con particular intensidad en las publicaciones de actos corteses valorizantes de Facebook, donde los internautas se muestran muy cariñosos con sus lazos más cercanos. Esta preponderancia se advierte especialmente en publicaciones relacionadas con logros personales, cumpleaños, aniversarios o cambios de imagen, mientras que en temas más polémicos, como política o debates sociales, los comentarios tienden a ser más polarizados y pueden incluir expresiones de descortesía. Además, esta tendencia afectuosa se vincula con el contexto de la pandemia y la pospandemia, períodos durante los cuales las interacciones sociales se convirtieron en un medio esencial para mantener la conexión afectiva y la cercanía social. En (8), el autor comparte un recuerdo junto a su hija, y un amigo comenta elogiándolos y manifestándoles su afecto mediante un *sticker* de una niña besando a su gato.

(8) Comentador (recordar): Guapísimos

Por lo que respecta a la función ilocutiva, de manera similar a los emoticonos, los emojis y los *kaomojis*, los *stickers* pueden indicar la fuerza ilocutiva del enunciado al que acompañan, facilitando así la correcta interpretación del mensaje y de la intención del emisor (Dresner y Herring 2010: 255-256). Como se expondrá en el apartado 3.3, los *stickers* pueden transmitir cortesía y utilizarse como estrategias para intensificar la cortesía valorizante en los actos de habla expresivos. Igualmente, pueden servir para señalar el tono humorístico, irónico o sarcástico de los mensajes. En nuestros datos, únicamente se localizaron dos *stickers* de risa, los cuales marcan el tono humorístico y generan complicidad con el otro, como en (9), donde una comentadora responde a otra con un *sticker* de Snoopy riendo, lo que potencia el tono jocoso de su intercambio.

(9) Comentadora (felicitar cumpleaños): Un día cuando celebremos todo lo pendiente, te voy a contar una cosa

Asimismo, dentro de la función ilocutiva, otro uso de los *stickers* es realizar actos de habla de manera autónoma. En los comentarios del corpus se han hallado *stickers* utilizados para felicitar, expresar buenos deseos, agradecer, elogiar, saludar, mostrar acuerdo, indicar aprobación, manifestar apoyo o responder a un agradecimiento con fórmulas de cortesía (por ejemplo, *de nada*)[10]. La mayoría de estos actos son expresivos, si bien mostrar acuerdo se clasifica como acto representativo (asertivo), ya que compromete al hablante con la verdad del contenido expresado (Searle 1976: 10). En (10), el autor reitera en una segunda respuesta su agradecimiento a los seguidores por las felicitaciones recibidas por su aniversario de bodas mediante un *sticker* de agradecimiento en inglés.

(10) Autor (felicitar aniversario): Muchas gracias a tod@s!!! 🥰🥰

Autor (felicitar aniversario):

Otra de las funciones comunicativas de los *stickers* es su papel en la gestión interaccional, ya que favorecen y estimulan las interacciones sociales.

---

[10] Aunque un comentador empleó un *sticker* con un corazón y el mensaje *SORRY Hola Te Amo Adiós*, que contenía una disculpa y una despedida, no se consideraron estos actos de habla, ya que, en el contexto comunicativo, el comentador eligió dicho *sticker* particularmente por el símbolo de corazón y con la intención de mostrar afecto.

A diferencia de la naturaleza sincrónica de los mensajes instantáneos, en Facebook las interacciones generadas en los comentarios como respuesta a las publicaciones de los autores suelen ser asincrónicas y se caracterizan por carecer de un cierre definido. Cualquier usuario puede retomar la conversación en un futuro, dado que esta permanece grabada en la plataforma. Por eso, no es común que aparezcan saludos y menos aún fórmulas de despedida en los comentarios. En (11), un amigo comenta los buenos deseos navideños de la autora, primero con un *sticker* de saludo y luego elogia su belleza. También son poco frecuentes los *stickers* que indican la escucha activa, los cuales permiten mostrar atención, reconocimiento y seguimiento. En (12), la autora comparte un recuerdo con una reflexión sobre su amor por el teatro y una amiga la anima y le muestra con emojis y con el *sticker* del pulgar hacia arriba su atención, apoyo y validación.

(11) Comentador (desear Feliz Año Nuevo):

Comentador: Que guapísima esta

(12) Comentadora (recordar): A por ello.. 👍👍
Comentadora:

Por otro lado, los *stickers*, en consonancia con los emojis, emoticonos y *kaomojis*, pueden funcionar como gestos digitales (Gawne y McCulloch 2019; McCulloch 2019: 232), asumiendo una función gestual que transmite significados a través de representaciones visuales. Sin embargo, su mayor complejidad y expresividad favorece que representen el lenguaje corporal y las expresiones faciales de forma más explícita y exhaustiva que otros tipos de graficones (Konrad *et al.* 2020: 222; Oberwinkler 2023: 256; Sampietro 2023b: 50). Junto a los gestos de saludo, en el corpus se han identificado *stickers* de gestos que expresan emociones, así como gestos que Gawne y McCulloch (2019) llaman de retroalimentación, ya que sirven como señales de escucha hacia el emisor y como reacciones a su intervención. Además, muestran reconocimiento o sintonía con el tema o el sentimiento expresado por el interlocutor. En concreto, se han localizado gestos de aprobación, acuerdo, aplauso, apoyo, brindis y celebración. En (13), una amiga responde a la felicitación de logro de la autora, dándole la enhorabuena junto a un *sticker* de aplauso.

(13) Comentadora (felicitar logro): Enhorabuena M.

Igualmente, los *stickers* pueden llevar a cabo una función estética cuando se usan para embellecer los mensajes, de manera que llaman la atención, transmiten una actitud positiva y enriquecen la experiencia comunicativa (Oberwinkler 2023: 256). Estos *stickers* no califican el texto, sino que lo adornan y, en muchos casos, aportan información redundante (Yus Ramos 2022: 96). En (14), el autor agradece las felicitaciones de cumpleaños recibidas, y un amigo lo felicita añadiendo a su mensaje un *sticker* de un hombre con una tarta y confeti. Este *sticker* añade colorido al mensaje, pero no aporta información nueva. De hecho, las felicitaciones de cumpleaños suelen contener emojis y *stickers* de tartas, confetis, globos, caras festivas o botellas descorchadas.

(14) Comentador (agradecer felicitaciones de cumpleaños): Muchas felicidades.. Un abrazo

Por último, la función representativa o sustitutiva se aplica a aquellos *stickers* cuya naturaleza icónica se emplea con el fin de ejercer como referentes visuales del contenido, ya sea al sustituir literalmente una palabra o idea, o al servir como representaciones metafóricas. En (15), la amiga trata de animar al autor, que se encuentra enfermo, con un *sticker* que equivale a la interjección «ánimo» (Vela Delfa y Cantamutto 2021: 73). Asimismo, son muy abundantes los *stickers* con valor metafórico, como los corazones, las rosas o los jarrones de flores, que expresan afecto hacia el interlocutor, tal como se observa en (16), donde un amigo manifiesta su cariño al autor.

(15) Comentadora (agradecer ayuda): Te deseo que te encuentres bien A. Comentadora:

(16) Comentador (recordar):

### 3.3 *La cortesía valorizante y el uso de* stickers

Las felicitaciones, los buenos deseos, los agradecimientos, las muestras de sentimientos positivos, las expresiones de condolencias y los recuerdos compilados en el corpus de Facebook son actos expresivos de cortesía valorizante que los autores comparten para ensalzar la imagen de los destinatarios de sus publicaciones. Del mismo modo, las publicaciones utilizadas por los autores para presumir de su propia imagen y recibir elogios permiten analizar los cumplidos como actos corteses valorizantes. Todos estos actos halagadores de la imagen suelen recibir respuestas de los comentadores también orientadas a la cortesía valorizante, principalmente en forma de afecto, elogios, felicitaciones, buenos deseos y agradecimientos.

A diferencia de otras redes, como Twitter, que tienden a organizarse en torno al debate público y la interacción con desconocidos, Facebook se configura con frecuencia como una red familiar, en la que predominan el mantenimiento y el refuerzo de relaciones sociales preexistentes fuera de línea (Ellison *et al.* 2007: 1144). Por eso, como señala Hernández Flores (2002: 67) con la conversación presencial española, las interacciones entre autores y comentadores en Facebook equilibran ambas imágenes sociales, y fortalecen los vínculos de parentesco, amistad y compañerismo. En este ecosistema comunicativo, los *stickers* pueden funcionar como estrategias de cortesía valorizante, al acompañar e intensificar los mensajes escritos, pero también pueden actuar de manera autónoma como actos corteses. Es importante señalar que en el corpus analizado solo se han localizado *stickers* que intensifican la cortesía valorizante, sin que se haya registrado ninguno que cumpla funciones de atenuación ni de descortesía. Este predominio puede explicarse, al menos en parte, por el tipo de publicaciones analizadas (felicitaciones, buenos deseos, agradecimientos, cambios de imagen de perfil, etc.), que favorecen la expresión de cortesía.

Los *stickers* se utilizan como mecanismos de intensificación de la cortesía valorizante expresada en los comentarios (Ikuta 2022: 61). Al igual que los emojis, su uso favorece que los mensajes transmitan y refuercen la cortesía (Padilla 2024: 55). En (17), por ejemplo, una amiga felicita a la autora por su cumpleaños y le expresa su afecto, intensificando su mensaje con un *sticker* amoroso de Angry Birds. Tal como sostienen Brown y Gilman (1989: 159), el afecto incrementa la cortesía, mientras que su disminución la reduce. En el corpus se evidencia una alta frecuencia de *stickers* afectuosos que operan como recursos visuales para intensificar la cortesía valorizante, enaltecer aún más la imagen de los destinatarios y, simultáneamente, reforzar la propia imagen de los emisores.

(17) Comentadora (agradecer felicitaciones de cumpleaños): Felicidades muchos besos!!!!!!
Comentadora:

Análogamente, los *stickers* pueden transmitir cortesía valorizante por sí mismos. Así, en (18), el autor comparte un recuerdo de las medallas obtenidas en un campeonato de atletismo, y un amigo lo elogia con el *sticker SUPER!*, con el que ensalza su imagen y manifiesta alegría por su éxito.

(18) Comentador (recordar):

La Tabla 4 presenta la frecuencia de *stickers* según los actos de habla principales de las publicaciones analizadas, así como el género de los participantes, diferenciando entre autores y comentadores.

TABLA 4. *Frecuencia de stickers por acto de habla y género*

| Acto | Autores | | Comentadores | | Total | | Suma |
|---|---|---|---|---|---|---|---|
| | H. | M. | H. | M. | H. | M. | |
| Felicitación | 1 | 3 | 9 | 56 | 10 | 59 | 69 |
| Deseo | 2 | 6 | 8 | 71 | 10 | 77 | 87 |
| Agradecimiento | 0 | 4 | 5 | 38 | 5 | 42 | 47 |
| Sentimiento + | 0 | 1 | 7 | 24 | 7 | 25 | 32 |
| Condolencias | 0 | 0 | 0 | 5 | 0 | 5 | 5 |
| Recuerdos | 0 | 1 | 10 | 46 | 10 | 47 | 57 |
| Presumir | 0 | 1 | 5 | 35 | 5 | 36 | 41 |
| **Total** | 3 | 16 | 44 | 275 | 47 | 291 | 338 |

Como se puede observar, las publicaciones que expresan buenos deseos (87) y las felicitaciones (69) son los actos de cortesía valorizante en los que autores y comentadores emplearon más *stickers* en los comentarios. Este patrón confirma que en Facebook los *stickers* favorecen de manera autónoma la expresión de cortesía valorizante y, cuando acompañan a actos halagadores de la imagen, propician su intensificación. En contraste, en las expresiones de condolencias ningún autor utilizó *stickers*, y solo 5 mujeres comentadoras los añadieron, probablemente debido al carácter más delicado de este acto, que requiere un cuidado especial de la imagen del otro en situaciones de dolor.

En cuanto al género, como se mostró en el apartado 3.1, se comprobó que el 86,2 % (275) de los *stickers* fueron enviados por mujeres comentadoras y el 84,2 % (16) por mujeres autoras, quienes los emplean como recursos de cortesía valorizante. Por acto de habla, tanto las mujeres comentadoras como las autoras usaron más *stickers* en los buenos deseos. En cambio, los hombres comentadores añadieron más *stickers* en los recuerdos, mientras que los autores lo hicieron principalmente en los buenos deseos. Asimismo, se aprecia una tendencia diferenciada en la elección de los *stickers* según el género. Los hombres tienden a compartir con mayor frecuencia *stickers* protagonizados por figuras humanas masculinas, en especial aquellos que muestran elogio, apoyo o camaradería, pero también los que incluyen muestras de afecto. El *sticker* más empleado por los participantes masculinos es el de Daily Duncan, que elogia con el número «10» 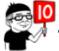. Por su parte, dado que las mujeres suelen ser más expresivas e interactivas en entornos digitales, los *stickers* que utilizan, como ya se ha señalado en otras investigaciones (Y. Kato 2017; Oberwinkler 2023: 256; Yang *et al.* 2023: 835), se caracterizan por un tono más tierno y afectuoso, e incorporan tanto figuras humanas como animales, de apariencia masculina o femenina. El más compartido en este grupo es el *sticker* amoroso de Hacker girl 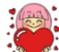.

4. CONCLUSIONES

El análisis del corpus de publicaciones y comentarios producidos durante y después de la pandemia de la COVID-19 en Facebook permite extraer varias conclusiones sobre el uso comunicativo de los *stickers* en contextos sociales de interacción digital. En un momento tan vulnerable y complejo, los *stickers* contribuyeron a la cohesión social en la comunicación mediada por ordenador. En primer lugar, entre los comentadores predominan los *stickers* desnudos frente a los que acompañan mensajes de texto, aunque entre los autores no se percibe una diferencia tan marcada. La mayor parte de estos graficones carece de texto, y los que los incluyen aparecen mayoritariamente en inglés y en mayúsculas. Esto contrasta con los resultados de Oberwinkler (2023: 246) sobre el empleo de *stickers* por usuarios japoneses en la aplicación de mensajería LINE, donde los *stickers* textuales eran mayoría. De este modo, en el ámbito español, su presencia en Facebook sigue siendo todavía reducida. Temáticamente, sobresalen los *stickers* que representan emociones y gestos humanos, así como aquellos relacionados con festividades y celebraciones.

En segundo lugar, los *stickers* en Facebook cumplen múltiples funciones comunicativas, lo que confirma su carácter multifuncional. Se delimitaron seis funciones principales: afectiva, ilocutiva, interaccional, gestual, estética y representativa o sustitutiva. La afectiva, subdividida en emotiva y afectuosa, es la más frecuente, lo que coincide con la imagen afiliativa española (Hernández Flores 2002: 92). La función ilocutiva también es relevante, ya que los *stickers* posibilitan indicar la fuerza ilocutiva, marcar el tono, atenuar o intensificar mensajes, transmitir cortesía o realizar actos de habla autónomos. Además, facilitan la gestión interaccional, embellecen el mensaje y pueden actuar como gestos, o sustituir palabras o metáforas.

En tercer lugar, los *stickers* desempeñan un papel activo en la expresión de la cortesía valorizante. No solo intensifican los actos halagadores de la imagen, sino que también pueden expresar cortesía por sí mismos. Los

*stickers* afectuosos refuerzan la cortesía valorizante, enaltecen la imagen de los destinatarios y fortalecen los vínculos sociales, un efecto especialmente significativo durante la situación de aislamiento social vivida en la pandemia.

Finalmente, desde un punto de vista sociopragmático, se constatan diferencias de género en su uso. En el corpus analizado en español, se identifican patrones en gran medida coincidentes con los descritos en investigaciones previas sobre *stickers* en contextos asiáticos, en aplicaciones como LINE y WeChat (Y. Kato 2017; Oberwinkler 2023: 256; Yang *et al.* 2023: 835): las mujeres, autoras y comentadoras, compartieron la mayor proporción de *stickers*, sobre todo en expresiones de buenos deseos, y utilizaron representaciones humanas y animales de aspecto tierno y afectuoso. Por su parte, los hombres enviaron menos *stickers*, pero centrados en figuras humanas masculinas. Además, los comentadores agregaron más *stickers* en publicaciones de recuerdos, mientras que los autores lo hicieron particularmente en manifestaciones de buenos deseos.

En síntesis, los *stickers* se consolidan como recursos multimodales, expresivos y multifuncionales habituales en las interacciones digitales cotidianas de Facebook, facilitando la cortesía valorizante y realzando las imágenes sociales. Su utilización en español articula formas de expresividad documentadas en contextos asiáticos con características propias de la cultura de acercamiento española y de la macroplataforma analizada. La tendencia apunta a un incremento en su uso, e incluso a una integración progresiva de emojis y *stickers* como un mismo graficón (Herring 2018: 14; Konrad *et al.* 2020: 223). Además, algunos *stickers* incorporan rasgos de otros recursos multimodales, como los memes (Sampietro 2023b: 13), lo que abre nuevas líneas para el estudio de esta hibridación multimodal en contextos digitales.

Por último, futuras investigaciones podrían abordar, por un lado, un estudio diacrónico del uso de los *stickers* en Facebook, con el fin de observar su evolución a lo largo del tiempo. Por otro lado, dado que el corpus analizado incluye el período de la pandemia y la pospandemia de la COVID-19, resulta conveniente realizar estudios comparativos que examinen las diferencias en su uso, tipología y funciones comunicativas en etapas prepandémicas, pandémicas y pospandémicas. De igual forma, sería de interés ampliar el análisis a otras plataformas digitales y considerar otras variables, como la edad y la orientación sexual, así como las diferencias interculturales y transculturales en su empleo.

# BIBLIOGRAFÍA

ADAMI, Elisabetta y Carey JEWITT (2016): «Special Issue: social media and the visual», *Visual Communication* 15, 263-270 (DOI: https://doi.org/10.1177/1470357216644153).

ALBELDA MARCO, Marta (2003): «Los actos de refuerzo de la imagen en la cortesía peninsular». En Diana Bravo (ed.), *Actas del primer coloquio del Programa EDICE: La perspectiva no etcentrista de la cortesía: identidad sociocultural de las comunidades hispanohablantes*, Estocolmo: Stockholms Universitet, 298-305.

ALVARADO ORTEGA, M. Belén y Esther LINARES BERNABÉU (2024): «Análisis pragmático de los stickers humorísticos en los grupos de WhatsApp», *Pragmalingüística* 33, 11-44 (DOI: https://doi.org/10.25267/Pragmalinguistica.2024.i32.01).

BAE, Gahyeon, Daehyun KWAK y Youn-Kyung LIM (2024): «Why I choose this sticker when chatting with you: exploring design considerations for sticker recommendation services in mobile instant messengers», *Proceedings of the ACM on Human-Computer Interaction* 8, 1-28 (DOI: https://doi.org/10.1145/3687063).

BATEMAN, John A. (2008): *Multimodality and genre: a foundation for the systematic analysis of multimodal documents*, Londres: Palgrave Macmillan (DOI: https://doi.org/10.1057/9780230582323_5).

BATEMAN, John A., Janina WILDFEUER y Tuomo HIIPPALA (2017): *Multimodality: foundations, research and analysis. A problem-oriented introduction*, Berlín: Mouton de Gruyter (DOI: https://doi.org/10.1515/9783110479898).

BEIßWENGER, Michael y Angelika STORRER (2008): «Corpora of computer-mediated communication». En Anke Lüdeling y Merka Kytö (eds.), *Corpus linguistics. An international handbook*, Berlín: Mouton de Gruyter, 292-308.

BERNAL LINNERSAND, María (2007): *Categorización sociopragmática de la cortesía y de la descortesía: un estudio de la conversación coloquial española*, Tesis doctoral de la Stockholms Universitet, Estocolmo.

BRAVO, Diana (1999): «¿Imagen "positiva" vs. imagen "negativa"?: pragmática socio-cultural y componentes de face», *Oralia: Análisis del Discurso Oral* 2, 155-184 (DOI: https://doi.org/10.25115/oralia.v2i.8533).

BRAVO, Diana (2002): «Actos asertivos y cortesía: imagen del rol en el discurso de académicos argentinos». En María E. Placencia y Diana Bravo (eds.), *Actos de habla y cortesía en español*, Múnich: Lincom Europa, 141-174.

BROWN, Penelope y Stephen C. LEVINSON (1987): *Politeness: some universals in language usage*, Cambridge: Cambridge University Press (DOI: https://doi.org/10.1017/CBO9780511813085).

BROWN, Roger y Albert GILMAN (1989): «Politeness theory and Shakespeare's four major tragedies», *Language in Society* 18, 159-212 (DOI: https://doi.org/10.1017/S0047404500013464).

CANTAMUTTO, Lucía (2023): «Estilo digital, lenguaje juvenil y gestión de vínculos: del lenguaje SMS al modo sticker», *Revista Argentina de Investigación Educativa* 3, 199-225. <https://ri.conicet.gov.ar/handle/11336/224499>.

CARRASCO SANTANA, Antonio (1999): «Revisión y evaluación del modelo de cortesía de Brown & Levinson», *Pragmalingüística* 7, 1-44 (DOI: https://doi.org/10.25267/Pragmalinguistica.1999.i7.01). <https://revistas.uca.es/index.php/pragma/article/view/499>.

CHEN, Zhenpeng, Xuan LU, Wei AI, Huoran LI, Qiaozhu MEI y Zuanzhe LIU (2018): «Through a gender lens: learning usage patterns of emojis from large-scale android users», *Proceedings of the 2018 World Wide Web Conference (WWW '18)*, Ginebra, 763-772 (DOI: https://doi.org/10.1145/3178876.3186157).

CRYSTAL, David (2011): *Internet linguistics: a student guide*, Londres: Routledge (DOI: https://doi.org/10.4324/9780203830901).

DAINAS, Ashley y Susan C. HERRING (2021): «Interpreting emoji pragmatics». En Chaoqun Xie, Francisco Yus Ramos y Hartmut Haberland (eds.), *Approaches to internet pragmatics: theory and practice*, Ámsterdam: John Benjamins Publishing Company, 107-144 (DOI: https://doi.org/10.1075/pbns.318.04dai).

DATAREPORTAL (2025): «Digital 2025 Global Overview». <https://datareportal.com/global-digital-overview>.

DE SETA, Gabriele (2018): «Biaoqing: the circulation of emoticons, emoji, stickers, and custom images on Chinese digital media plataforms», *First Monday* 23 (DOI: http://dx.doi.org/10.5210/fm.v23i9.9391).

DERKS, Daantje, Arjan BOS y Jasper V. GRUMBKOW (2007): «Emoticons and social interaction on the Internet: the importance of social context», *Computers in Human Behavior* 23, 842-849 (DOI: https://doi.org/10.1016/j.chb.2004.11.013).

DERKS, Daantje, Arjan BOS y Jasper V. GRUMBKOW (2008): «Emoticons and online message interpretation», *Social Science Computer Review* 26, 379-388 (DOI: https://doi.org/10.1177/0894439307311611).

DIMOCK, Michael (2019): «Defining generations: where Millenials end and Generation Z begins», Pew Research Center. <https://www.pewresearch.org/fact-tank/2019/01/17/where-millennials-end-and-generation-z-begins/>.

DRESNER, Eli y Susan C. HERRING (2010): «Functions of the non-verbal in CMC: emoticons and illocutionary force», *Communication Theory* 20, 249-362 (DOI: https://doi.org/10.1111/j.1468-2885.2010.01362.x).

ELLISON, Nicole B., Charles STEINFIELD y Cliff LAMPE (2007): «The benefits of facebook "friends": social capital and college students' use of online social network sites», *Journal of Computer-Mediated Communication* 12, 1143-1168 (DOI: https://doi.org/10.1111/j.1083-6101.2007.00367.x).

ESCANDELL VIDAL, M. Victoria (1996): *Introducción a la pragmática*, Barcelona: Ariel.

FUNDÉU (2017): «Sticker, alternativas en español». <https://www.fundeu.es/recomendacion/sticker-alternativas-en-espanol/>.

GAWNE, Lauren y Gretchen MCCULLOCH (2019): «Emoji as digital gestures», *Language@internet* 17. <https://scholarworks.iu.edu/journals/index.php/li/article/view/37786>.

GOFFMAN, Ervin (1967): *Interactional ritual: essays on face-to-face behavior*, Nueva York: Garden City.

GUMPERZ, John J. (1982): *Discourse strategies*, Cambridge: Cambridge University Press (DOI: https://doi.org/10.1017/CBO9780511611834).

HALTÉ, Pierre (2019): «Emojis, émoticônes, smileys? Proposition de classement terminologique selon des critères sémiotiques et énonciatifs», *Interfaces numériques* 8, 365-386 (DOI: https://doi.org/10.25965/interfaces-numeriques.3956).

HASYM, Muhammad y Burhanuddin ARAFAH (2023): «Semiotic multimodality communication in the age of new media», *Studies in Media and Communication* 11, 96-103 (DOI: https://doi.org/10.11114/smc.v11i1.5865).

HAVERKATE, Henk (1994): *La cortesía verbal: estudio pragmalingüístico*, Madrid: Gredos.

HERNÁNDEZ FLORES, Nieves (2002): *La cortesía en la conversación española de familiares y amigos: la búsqueda de equilibrio entre la imagen del hablante y la imagen del destinatario*, Tesis doctoral de la Aalborg University, Dinamarca.

HERNÁNDEZ FLORES, Nieves (2013): «Actividad de imagen: caracterización y tipología en la interacción comunicativa», *Pragmática sociocultural: revista internacional sobre lingüística del español* 1, 175-198 (DOI: https://doi.org/10.1515/soprag-2012-0012).

HERRING, Susan C. (1996a): «Introduction». En Susan C. Herring (ed.), *Computer-mediated communication: linguistic, social, and cross-cultural perspectives*, Ámsterdam: John Benjamins Publishing Company, 1-12.

HERRING, Susan C. (1996b): «Linguistic and critical analysis of computer-mediated communication: some ethical and scholarly considerations», *The Information Society* 12, 153-168. (DOI: https://doi.org/10.1080/019722496129576).

HERRING, Susan C. (1999): «Interactional coherence in CMC», *Journal of Computer-Mediated Communication* 4 (DOI: https://doi.org/10.1111/j.1083-6101.1999.tb00106.x).

HERRING, Susan C. (2018): «Emergent forms of computer-mediated communication and their global implications», *LinguaPax Review 2017*. <https://homes.luddy.indiana.edu/herring/linguapax.pdf>.

HERRING, Susan C. y Ashley DAINAS (2017): «"Nice picture comment!": graphicons in Facebook comment threads», *Proceedings of the Fiftieth Hawaii International Conference on System Sciences*, 2185-2194 (DOI: https://doi.org/10.24251/HICSS.2017.264).

HERRING, Susan C. y Ashley DAINAS (2018): «Receiver interpretations of emojis functions: a gender perspective», *Proceedings of the 1st International Workshop on Emoji Understanding and Applications in Social Media (Emoji2018)*, 1-8. <https://api.semanticscholar.org/CorpusID:264712497>.

HERRING, Susan C. y Ashley DAINAS (2020): «Gender and age influences on interpretation of emoji functions», *ACM Transactions on Internet Technology* 3, 1-26 (DOI: https://doi.org/10.1145/3375629).

HERRING, Susan C. y Sharon STOERGER (2014): «Gender and (a)nonymity in computer-mediated communication». En Susan Ehrlich, Miriam Meyerhoff y Janet Holmes (eds.), *The handbook of language, gender, and sexuality*, Nueva York: John Wiley & Sons, 567-586 (DOI: https://doi.org/10.1002/9781118584248.ch29).

HINE, Christine (2000): *Virtual ethnography*, Thousand Oaks: SAGE Publications (DOI: https://doi.org/10.4135/9780857020277).

HOLMES, Janet (1995): *Women, men and politeness*, Londres. Routledge.

IKUTA, Shoko (2022): «On the linguistic norm of Japanese seniors in mobile texting interaction: as reflected in their sticker use and repair work», *Meiji Gakuin University, The Journal of English & American Literature and Linguistics* 137, 53-74.

JESSICA, Gitari y Elda FRANZIA (2017): «The analysis of LINE sticker character "Cony special edition"», *Humaniora* 8, 291-301 (DOI: https://doi.org/10.21512/humaniora.v8i3.3904).

JEWITT, Carey, Jeff BEZEMER y Kay O'HALLORAN (2016): *Introducing multimodality*, Londres: Routledge (DOI: https://doi.org/10.4324/9781315638027).

JEZOUIT, Britanny (2017): «Are stickers the new emoji?», Envato Blog. <https://elements.envato.com/learn/stickers-new-emoji>.

KANEYASU, Michiko (2022): «Multimodal strategies for balancing formality and informality: the role of kaomoji in online comment-reply interactions», *Internet Pragmatics* 5, 143-164 (DOI: https://doi.org/10.1075/ip.00071.kan).

KANG, Hyunmin, YounJung PARK, Yonghwan SHIN, Hobin CHOI y Sungtae KIM (2022): «What makes consumers' intention to purchase paid stickers in personal Messenger? The role of personality and motivational factors», *Frontiers in Psychology* 12, 1-19 (DOI: https://doi.org/10.3389/fpsyg.2021.678803).

KATO, Shogo y Yuuki KATO (2018): «Exploring potential factors in sticker use among Japanese young adults: effects of gender and text messaging dependency», *International Journal of Virtual Communities and Social Networking* 10, 1-23 (DOI: https://doi.org/10.4018/IJVCSN.2018040101).

KATO, Yuuki (2017): «Basic survey on the use of LINE stamps», *Journal of Information and Media Studies* 3, 21-34.

KATSUNO, Hirofumi y Christine YANO (2002): «Face to face: on-line subjectivity in contemporary Japan», *Asian Studies Review* 26, 205-231 (DOI: https://doi.org/10.1080/10357820208713341).

KEMP, Simon (2025): «Digital 2025: Spain», DataReportal. <https://datareportal.com/reports/digital-2025-spain>.

KERBRAT-ORECCHIONI, Catherine (1992): *Les interactions verbales II*, París: Armand Colin (DOI: https://doi.org/10.1515/9783111678504-006).

KERBRAT-ORECCHIONI, Catherine (1996): *La conversation*, París: Seuil.

KIMBROUGH, Amanda M., Rosanna E. GUADAGNO, Nicole L. MUSCANELL y Janeann DILL (2013): «Gender differences in mediated communication: women connect more than do men», *Computers in Human Behavior* 29, 896-900 (DOI: https://doi.org/10.1016/j.chb.2012.12.005).

KOCH, Peter y Wulf OESTERREICHER (2007): *Lengua hablada en la Romania: español, francés, italiano*, Madrid: Gredos.

KOCH, Timo K., Peter ROMERO y Clemens STACHL (2022): «Age and gender in language, emoji, and emoticon usage in instant messages», *Computers in Human Behavior* 126, 1-12 (DOI: https://doi.org/10.1016/j.chb.2021.106990).

KONRAD, Artie, Susan C. HERRING y David CHOI (2020): «Sticker and emoji use in Facebook Messenger: implications for graphicon change», *Journal of Computer-Mediated Communication* 25, 217-235 (DOI: https://doi.org/10.1093/jcmc/zmaa003).

KRESS, Gunther y Theo VAN LEEUWEN (2001): *Multimodal discourse. The models and media of contemporary communication*, Londres: Arnold.

LEE, Joon Y., Nahi HONG, Soomin KIM, Jonghwan OH y Joonhwan LEE (2016): «Smiley face: why we use emoticon stickers in mobile messaging», *Proceedings of the 18th International Conference on Human-Computer*

*Interaction with Mobile Devices and Services Adjunct (MobileHCI '16)*, 760-766 (DOI: https://doi.org/10.1145/2957265.2961858).

LEE, Wen-Hsuan y Yu-Hsun LIN (2019): «Online communication of visual information: stickers' functions of self-expression and conspicuousness», *Online Information Review* 44, 43-61 (DOI: https://doi.org/10.1108/OIR-08-2018-0235).

LIU, Siying y Renji SUN (2020): «To express or to end? Personality traits are associated with the reasons and patterns for using emojis and stickers», *Frontiers in Psychology* 11, 1-11. (DOI: https://doi.org/10.3389/fpsyg.2020.01076).

LÓPEZ MORALES, Humberto (1994): *Métodos de investigación lingüística*, Salamanca: Ediciones Colegio de España.

MA, Xiaojuan (2016): «From Internet memes to emoticon engineering: insights from the Baozou comic phenomenon in China». En Maasaki Kurosu (ed.), *Human-Computer Interaction. Novel User Experiences. HCI 2016. Lecture notes in computer science*, Cham: Springer, 15-27 (DOI: https://doi.org/10.1007/978-3-319-39513-5_2).

MCCULLOCH, Gretchen (2019): *Because internet: understanding the new rules of language*, Nueva York: Riverhead Books.

MORENO, Manuel (2025): «Cómo dibujar en los mensajes directos de Instagram», TreceBits. <https://www.trecebits.com/como-dibujar-mensajes-directos-instagram/>.

MOSCHINI, Ilaria (2016): «The "face with tears of joy" emoji: a socio-semiotic and multimodal insight into a Japan-America mash-up», *HERMES-Journal of Language and Communication in Business* 55, 11-25 (DOI: https://doi.org/10.7146/hjlcb.v0i55.24286).

OBERWINKLER, Michaela (2023): «Digital stickers in Japanese LINE communication», *IMAGE. Zeitschrift für interdisziplinäre Bildwissenschaft* 38, 238-262 (DOI: http://dx.doi.org/10.25969/mediarep/22339).

PADILLA, Xose A. (2024): «Los emojis en WhatsApp: funciones pragmático-discursivas y multimodalidad», *Oralia: Análisis del Discurso Oral* 27, 49-72 (DOI: https://doi.org/10.25115/oralia.v27i1.8630).

PÉREZ-SABATER, Carmen (2019): «Emoticons in relational writing practices on WhatsApp: some reflections on gender». En Patricia Bou-Franch y Pilar Garcés-Conejos Blitvich (eds*.), Analyzing digital discourse: new insights and future directions*, Londres: Palgrave Macmillan, 163-189 (DOI: https://doi.org/10.1007/978-3-319-92663-6_6).

PÉREZ-SABATER, Carmen (2025): «"Emojis are grown-up stuff": analysing graphical elements in digital relational contexts», *Internet Pragmatics* 8, 152-185 (DOI: https://doi.org/10.1075/ip.00125.per).

PLACENCIA, María Elena y Zohreh R. ESLAMI (2020): *Complimenting behavior and (self-)praise across social media new contexts and new*

*insights*, Ámsterdam: John Benjamins Publishing Company (DOI: https://doi.org/10.1075/pbns.313).

PORRINO-MOSCOSO, Laura M. (2024): *Sociopragmática en el universo de Facebook e Instagram: la cortesía retrogratificadora y la afectuosidad conversacional de los usuarios ordinarios y de los personajes influyentes*, Tesis doctoral de la Universidad de Salamanca, España.

PROVINE, Robert R., Robert J. SPENCER y Darcy L. MANDELL (2007): «Emotional expression online emoticons punctuate website text messages», *Journal of Language and Social Psychology* 26, 299-307. (DOI: https://doi.org/10.1177/0261927X06303481).

ROJO SÁNCHEZ, Guillermo (2021): *Introducción a la lingüística de corpus en español*, Londres: Routledge (DOI: https://doi.org/10.4324/9781003119760).

SADIA, Hina y Muhammad S. HUSSAIN (2023): «Use of emojis and stickers for online interaction facilitation: a gender-based semiotic discourse analysis», *Global Digital & Print Media Review* 6, 109-128 (DOI: https://doi.org/0.31703/gdpmr.2023(VI-II).09).

SAMPIETRO, Agnese (2023a): «El auge de los "stickers" en WhatsApp y la evolución de la comunicación digital», *Círculo de Lingüística Aplicada a la Comunicación* 94, 271-285 (DOI: https://doi.org/10.5209/clac.83860).

SAMPIETRO, Agnese (2023b): *Lengua e imagen en la comunicación digital*, Madrid: Arco/Libros.

SEARLE, John R. (1976): «A classification of illocutionary acts», *Language in Society* 5, 1-23 (DOI: https://doi.org/10.1017/S0047404500006837).

STEINBERG, Marc (2020): «LINE as super app: platformization in East Asia», *Social Media + Society*, 1-10 (DOI: https://doi.org/10.1177/2056305120933285).

SUSANTO, Billy S. (2018): «The evolution of communication and why stickers matter», Forbes. <https://www.forbes.com/sites/forbestechcouncil/2018/07/31/the-evolution-of-communication-and-why-stickers-matter/?sh=5dbb15612b2f>.

TANG, Ying y Khe F. HEW (2018): «Emoticon, emoji, and sticker use in computer-mediated communications: understanding its communicative function, impact, user behavior, and motive». En Liping Deng, W. Will Ma y Cheuk Fong (eds.), *New media for educational change. Educational communications and technology yearbook*, Singapur: Springer, 191-201 (DOI: https://doi.org/10.1007/978-981-10-8896-4_16).

TANG, Ying y Khe F. HEW (2019): «Emoticon, emoji, and sticker use in computer-mediated communication: a review of theories and research findings», *International Journal of Communication* 13, 2457-2483. <https://ijoc.org/index.php/ijoc/article/view/10966>.

TANG, Ying, Khe F. HEW, Susan C. HERRING y Quian CHEN (2021): «(Mis)communication through stickers in online group discussions: a multiple-case study», *Discourse & Communication* 15, 1-25 (DOI: https://doi.org/10.1177/17504813211017707).

TOSSELL, Chad C., Phillip KORTUM, Clayton SHEPARD, Laura BARG-WALKOW, Ahmad RAHMATI y Lin ZHONG (2012): «A longitudinal study of emoticon use in text messaging from smartphones», *Computers in Human Behavior* 28, 659-663 (DOI: https://doi.org/10.1016/j.chb.2011.11.012).

UNICODE (2024): «About emoji», Unicode. <https://home.unicode.org/emoji/about-emoji/>.

VELA DELFA, Cristina y Lucía CANTAMUTTO (2021): *Los emojis en la interacción digital escrita*, Madrid: Arco/Libros.

VELA DELFA, Cristina y Lucía CANTAMUTTO (2024): «Del emoticono a sticker: tendencias en el uso de graficones en interacciones de WhatsApp en lengua española», *Lengua y Sociedad. Revista de Lingüística Teórica y Aplicada* 23, 777-790 (DOI: https://doi.org/10.15381/lengsoc.v23i2.27113).

VELA DELFA, Cristina y Lucía CANTAMUTTO (2025): «Emojis y otros graficones». En M. Elena Placencia y Alejandro Parini (eds.), *Introducción al estudio del discurso digital en español*, Londres: Routledge, 175-189 (DOI: https://doi.org/10.4324/9781003327097-15).

WIERZBICKA, Anna (1991): *Cross-cultural pragmatics: the semantics of human interaction*, Berlín: Mouton de Gruyter (DOI: https://doi.org/10.1515/9783112329764).

YANG, Dongdong, Laura LABATO y Shardé M. DAVIS (2023): «Cuteness in mobile messaging: an exploration of virtual "cute" sticker use in China and the United States», *International Journal of Communication* 17, 819-840. <https://ijoc.org/index.php/ijoc/article/view/19253/4026>.

YUS RAMOS, Francisco (2022): *Smartphone communication: interactions in the app ecosystem*, Londres: Routledge (DOI: https://doi.org/10.4324/9781003200574).

ZHOU, Rui, Jasmine HENTSCHEL y Neha KUMAR (2017). «Goodbye text, hello emoji: mobile communication on WeChat in China», *Proceedings of the 2017 CHI Conference on Human Factors in Computing Systems*, 748-759 (DOI: https://doi.org/10.1145/3025453.3025800).